\documentstyle[prl,aps]{revtex}
\input epsf

\tighten

\begin{document}


\title{A class of kinks in $SU(N)\times Z_2$}

\author{
Tanmay Vachaspati}
\address
{
Department of Astronomy and Astrophysics,
Tata Institute of Fundamental Research,\\
Homi Bhabha Road, 
Colaba, Mumbai 400005, India\\
and\\
Department of Physics,
Case Western Reserve University,\\
10900 Euclid Avenue,
Cleveland, OH 44106-7079, USA.}

\wideabs{
\maketitle

\begin{abstract}
\widetext
In a classical, quartic field theory with $SU(N) \times Z_2$ 
symmetry, a class of kink solutions can be found analytically for 
one special choice of parameters. We construct these solutions 
and determine their energies. In the limit 
$N\rightarrow \infty$, the energy of the kink is equal to that 
of a kink in a $Z_2$ model with the same mass parameter and 
quartic coupling (coefficient of ${\rm Tr}(\Phi^4)$). We prove 
the stability of the solutions to small perturbations but global 
stability remains unproven. We then argue that the continuum 
of choices for the boundary conditions leads to a whole 
space of kink solutions. The kinks in this space occur in classes 
that are determined by the chosen boundary conditions. Each class 
is described by the coset space $H/I$ where $H$ is the unbroken 
symmetry group and $I$ is the symmetry group that leaves the kink 
solution invariant. 
\end{abstract}
}

\narrowtext

\section{Introduction}
\label{introduction}

Classical solutions in field theories have been of
long-standing interest and have played an important
role in our understanding of a variety of phenomena.
The existence of solutions can be predicted on the 
basis of topology though the actual construction of 
the solutions is usually quite difficult and requires 
a certain amount of guesswork. Even after a solution
is found, one needs to check its perturbative stability.
And even after a solution is proved to be perturbatively 
stable, there is no guarantee that it is the globally 
stable solution. 

In this paper, we shall construct a class of kink solutions in 
purely scalar, $SU(N)\times Z_2$ field theories. The
analytic construction is made possible by a special choice
of parameters. We then prove the perturbative stability of the
kink solutions. As there is no known extension of Bogomolnyi's 
method \cite{Bog76} to this case, global stability will 
remain unproven. We will then address the question of whether there
might be other kink solutions in the model. We discuss the
possibility that various boundary conditions imposed at spatial 
infinity, while topologically equivalent, lead to different kink 
solutions. Each kink solution is continuously degenerate with a 
degeneracy determined by the invariance group of the solution.
We describe the problem of mapping the space of kink solutions.

The following few sections deal with the construction and perturbative
stability of the kink solutions that have been found. The space
of kinks is described in Sec. \ref{kinkspace} and this can be read
with minimal reference to the prior sections.

\section{The Model}
\label{themodel}

We will consider the 1+1-dimensional, classical field 
theory\footnote{The addition of gauge fields can easily
be accommodated in what follows as discussed in 
Sec. \ref{gaugefields}.}
\begin{equation}
L = {\rm Tr}[ (\partial _\mu \Phi )^2 ] - V(\Phi )
\label{lagrangian}
\end{equation}
\begin{equation}
V(\Phi ) = - m^2 {\rm Tr}[ \Phi ^2 ]+
      h ( {\rm Tr}[\Phi ^2  ])^2 +
      \lambda {\rm Tr}[\Phi ^4]  + V_0 
\label{lagrangianpotential}
\end{equation}
where $\Phi$ transforms in the adjoint representation of
$SU(N)$ ($N=2n+1$) and hence is an $N\times N$ hermitian matrix. 
$\Phi$ can also be written in terms of components:
\begin{equation}
\Phi = \sum_{a=1}^{N^2-1} \Phi^a T_a
\label{phicomponents}
\end{equation}
where $T_a$ are the (traceless) generators of $SU(N)$ and will 
be taken to be in the Gell-Mann representation \cite{Geo82} 
and normalized so that 
\begin{equation}
{\rm Tr} [T_a T_b] = {1\over 2} \delta_{ab} \ .
\label{tnormalization}
\end{equation}
The field components $\Phi^a$ are real. The constant $V_0$
is chosen so that the minimum of the potential is at $V=0$.
Note the absence of a cubic term in the model.

The center of $SU(N)$ is Z$_N$ and the group elements of the center
transform $\Phi$ by multiplication by factors of $\exp(i2\pi /N)$.
For even $N$, the transformation $\Phi \rightarrow -\Phi$ is 
included in the center of $SU(N)$. However, for odd $N$ it is not,
and the full symmetry of the model is $SU(N)\times Z_2$.
We will only be interested in odd values of $N$ since it is this
case that the model admits topological kink solutions.

We will consider parameters $h$ and $\lambda$ such that the
symmetry breaking is:
\begin{equation}
SU(N)\times Z_2 \rightarrow 
         {{SU(n+1)\times SU(n)\times U(1)} \over
            {Z_{n+1} \times Z_n }}
\label{symbreaking}
\end{equation}
where $N=2n+1$ ($n$ is a positive integer).
This symmetry breaking pattern is achieved if \cite{Lix74}
\begin{equation}
{h \over \lambda} > - {{N^2+3}\over {N(N^2-1)}} 
\label{paramconstraint}
\end{equation}
In fact, in the next section we will choose
\begin{equation}
{h\over \lambda} = - {3\over {N(N-1)}} \ .
\label{paramchoice}
\end{equation}
This choice is consistent with eq. (\ref{paramconstraint}) as
long as $N > 3$.

To fix the constant $V_0$, let the minimum of $V(\Phi )$ 
occur at
\begin{equation}
\Phi = \Phi_0 \equiv R \ {\rm diag} (p_1, p_2,...,p_N) 
\label{phivev}
\end{equation}
where $R$ is an overall normalization factor and the
matrix is normalized so that 
\begin{equation}
\sum_{i=1}^{N} p_i^2 = {1\over 2} \ .
\label{pinorm}
\end{equation}
Then the potential is extremized for
\begin{equation}
R = {{m} \over {\sqrt{\lambda '}}}
\label{Rvalue}
\end{equation}
where
\begin{equation}
\lambda ' = h + 4\lambda \sum_{i=1}^N p_i^4 =
h +  {{N^2+3}\over {N(N^2-1)}} \lambda  \ .
\label{lambdaprime}
\end{equation}
Then, $V(\Phi_0 )=0$ gives
\begin{equation}
V_0 = - {{m^4}\over {4\lambda '}} \ .
\label{V0}
\end{equation}
The condition that $\lambda ' > 0$ is precisely
the condition in eq. (\ref{paramconstraint}) 

We shall assume the parameters as constrained by 
eq. (\ref{paramconstraint}), in which case the unbroken
symmetry is given in eq. (\ref{symbreaking}) and 
the vacuum expectation value of $\Phi$ is
\begin{equation}
\Phi_0 = R \sqrt{2 \over {N(N^2-1)}}
            \pmatrix{n{\bf 1}_{n+1}&{\bf 0}\cr
                      {\bf 0}&-(n+1){\bf 1}_n\cr}
\label{phi0}
\end{equation}
where ${\bf 1}_n$ denotes the $n\times n$ unit matrix,
$N=2n+1$, $R$ is defined in eq. (\ref{Rvalue}).

\section{Kink solution}
\label{kinksolutionsection}

The model contains the breaking of a discrete $Z_2$
symmetry whenever $N$ is odd. Hence kink solutions
exist. Across these kink solutions, the vacuum expectation
value of $\Phi$ must be related by an element of the $Z_2$
group. Hence, if $\Phi_k$ denotes the kink solution,
$$
\Phi_k (x=-\infty ) = - U \Phi_k (x= +\infty ) U^{-1}
$$
where $U$ is an element of $SU(N)$.

Based on the experience with kinks in $SU(5)$ \cite{PogVac00},
we conjecture the following form for the kink solution:
\begin{equation}
\Phi_k (x) = f(x) {\bf M} + g(x) {\bf P} 
\label{phikconjecture}
\end{equation}
where $f(x)$ and $g(x)$ are unspecified functions as yet, 
and ${\bf M}$ and ${\bf P}$ are $SU(N)$ generators such that
\begin{equation}
{\rm Tr} ({\bf M}^2 ) = {1\over 2} = {\rm Tr} ({\bf P}^2 ) \\  
\label{MPproperties1}
\end{equation}
\begin{equation}
{\rm Tr} ({\bf MP} ) = 0 = {\rm Tr} ({\bf M}^3{\bf P} ) = 
                       {\rm Tr} ({\bf M}{\bf P}^3 ) \ .
\label{MPproperties2}
\end{equation}
Insertion of $\Phi_k$ in the potential gives
\begin{eqnarray}
V(\Phi_k) = 
&& - {{m^2}\over 2} (f^2 + g^2) +
              {{h}\over 4} (f^4 + g^4) + \nonumber \\
&&\lambda [\, f^4 {\rm Tr} ({\bf M}^4 ) + 
                 g^4{\rm Tr} ({\bf P}^4 )\, ] + \nonumber \\
&& [ \, {h\over 2} + 6 \lambda {\rm Tr} ( {\bf M}^2{\bf P}^2 )\, ] 
         f^2 g^2 - V_0 \ .
\label{vphik}
\end{eqnarray} 
The important realization that permits an analytic solution
to be found is that the cross-term containing both $f$ and
$g$ disappears if we choose
\begin{equation}
h = -12 \lambda {\rm Tr} ({\bf M}^2{\bf P}^2 ) \ .
\label{hcondition}
\end{equation}
In this case, the energy of the kink will separate into
two pieces, one depending only on $f$ and the other depending only
on $g$. So the energy will be a sum of two single field energies,
each of which can be treated directly or by Bogomolnyi's method.

Let us further choose the matrices ${\bf M}$ and ${\bf P}$
in such a way that the kink boundary conditions imply
\begin{equation}
f(+\infty ) = - f(-\infty ) \ , \ \ \ 
g(+\infty ) = + g(-\infty ) \ .
\label{fgbcs}
\end{equation}
Then, we find that $f$ and $g$ must satisfy
\begin{equation}
f'' + m^2 f + 
4\lambda [ 3{\rm Tr} ({\bf M}^2{\bf P}^2 ) - {\rm Tr} ({\bf M}^4 ) ] f^3 = 0
\label{fequation}
\end{equation}
\begin{equation}
g ^2 = - {{m^2} \over {4 \lambda }}
    {1\over { 3{\rm Tr} ({\bf M}^2{\bf P}^2 ) - {\rm Tr} ({\bf P}^4 )}} \ .
\label{gequation}
\end{equation}
The equation for $f$ will have the solution
\begin{equation}
f(x) = f_0 \tanh \left ( {m\over {\sqrt{2}}} x \right )
\label{ftanh}
\end{equation}
where 
\begin{equation}
f_0^2 = {{m^2}\over {4\lambda}}
         {1\over {{\rm Tr} ({\bf M}^4) - 3{\rm Tr} ({\bf M}^2{\bf P}^2 )}} \ .
\label{f0squared}
\end{equation}
The solution for $f$ is valid provided
\begin{equation} 
{\rm Tr} ({\bf M}^4 ) > 3{\rm Tr} ({\bf M}^2{\bf P}^2 )  
\label{condition1}
\end{equation}
and the (constant) solution for $g$ is valid provided
\begin{equation}
{\rm Tr} ({\bf P}^4 ) > 3{\rm Tr} ({\bf M}^2{\bf P}^2 )
\label{condition2}
\end{equation}

In addition to the properties in eqs. (\ref{MPproperties1})
and (\ref{MPproperties2}),
the conditions in eqs. (\ref{condition1}) and (\ref{condition2}),
${\bf M}$ and ${\bf P}$ must yield the desired vacuum expectation
values for $\Phi$ at spatial infinity. 
A choice of ${\bf M}$ and ${\bf P}$ that satisfies all
these conditions is:
\begin{eqnarray}
&{\bf M}&= \beta \pmatrix{ {\bf 1}_{N-1}&0\cr
                           0&-(N-1)\cr}\ ,  \nonumber \\  
&{\bf P}&= \sqrt{N} \beta \pmatrix{ {\bf 1}_{n}&0&0\cr
                                   0&-{\bf 1}_{n}&0\cr
                                   0&0&0\cr}
\label{MPchoice}
\end{eqnarray}
where,
\begin{equation}
\beta = {1\over {\sqrt{2N(N-1)}}} \ .
\label{betavalue}
\end{equation}
Straightforward calculation shows that all the conditions
on ${\bf M}$ and ${\bf P}$ are met provided that $N > 3$.

With this choice of ${\bf M}$ and ${\bf P}$, the kink solution
can now be written as:
\begin{equation}
\Phi_k (x) = {m\over {\sqrt{\lambda}}} \sqrt{{N-1}\over {N-3}}
   \left [ \tanh \left ( {m\over {\sqrt{2}}} x \right ) {\bf M}
             + \sqrt{N} {\bf P} \right ]  
\label{kinksolution}
\end{equation}
An alternative, more transparent, form of this solution is:
\begin{equation}
\Phi_k (x) =  \left ( {{1-F(x)}\over 2} \right ) \Phi_- +
           \left ( {{1+F(x)}\over 2} \right ) \Phi_+
\label{alternateform}
\end{equation}
where 
\begin{equation}
F(x) = \tanh \left ( {{m}\over {\sqrt{2}}} x \right ) \ ,
\label{Fdefn}
\end{equation}
and 
$\Phi_\pm \equiv \Phi(x=\pm \infty )$. Note that the alternate
form does not work for any chosen boundary condition. For 
example, if $\Phi_+ = -\Phi_-$, the form leads to the embedded
kink solution which is known to be unstable \cite{PogVac00}.

Now that we have shown that $\Phi_k$ is a solution within the
restricted ansatz in eq. (\ref{phikconjecture}), we also need
to show that it is a solution of the full theory. This is most
simply done by writing
\begin{equation}
\Phi = \Phi_k + \Psi = \Phi_k + \sum \psi_a T^a
\label{variation}
\end{equation}
and then checking that the energy density does not contain any 
terms that are linear in $\Psi$. In the quadratic terms in the
energy density, this will clearly by the case since $\Phi_k$
satisfies the equations of motion (\ref{fequation}) and 
(\ref{gequation}) and the generators satisfy the orthogonality
condition in eq. (\ref{tnormalization}). The only terms that
can potentially lead to a term linear in the $\psi_a$ come from
the quartic term, ${\rm Tr} (\Phi^4 )$, in the potential and are
of the form
$$
{\rm Tr} (\Phi_k^3 T^a) \psi^a
$$
For off-diagonal $T^a$, this vanishes since $\Phi_k^3$ is diagonal
and a product of a diagonal and an off-diagonal matrix is 
off-diagonal. For diagonal $T^a$, it is better to choose a
representation of the generators other than the Gell-Mann
representation. The choice of diagonal generators other
than ${\bf M}$ and ${\bf P}$ are:
\begin{eqnarray}
{\cal T}^i = \pmatrix{{\bf \tau}_i&{\bf 0}&0\cr
                         {\bf 0}&{\bf 0}_n&0\cr
                         0&0&0\cr}  \ \ (i=1,..,n-1)\\ 
{\cal T}^i = \pmatrix{{\bf 0}_n&{\bf 0}&0\cr
                        {\bf 0}&{\bf \tau}_{i-n+1}&0\cr
                        0&0&0\cr}  \ \ (i=n,...,2n-2) \nonumber \\ 
\label{newbasis}
\end{eqnarray}
where ${\bf \tau}_i$ ($i=1,..,n-1$) are the normalized, diagonal 
$SU(n)$ generators in the Gell-Mann representation. The off-diagonal
generators are the same as in the Gell-Mann basis.
It is easy to check that this basis is complete and the
generators satisfy the normalization in eq. (\ref{tnormalization}).

Now we are interested in checking if there are terms in the
potential that are linear in the components of $\Psi$ expanded
in the new basis. It is easy to check that
$$
{\rm Tr}( {\bf M}^3 {\cal T}^i ) = 0 = 
                  {\rm Tr} ({\rm P}^3 {\cal T}^i )
$$
Hence
$$
{\rm Tr}( \Phi_k^3 {\cal T}^i ) = 0
$$
and there are no linear terms in the components of $\Psi$
occurring in the energy density. The terms linear
in the perturbations along the generators ${\bf M}$ and
${\bf P}$ will vanish because these have already been chosen
to satisfy the equations of motion. 

Therefore there are no linear terms in $\Psi$ in the
energy density and $\Phi_k$ in eq. (\ref{kinksolution})
is indeed a solution.

\section{Kink energy}
\label{kinkenergy} 

The energy of the kink is:
\begin{equation}
E_k = \int dx [ {\rm Tr} (\Phi_k ')^2 + V(\Phi_k ) ]
\label{energyexpression}
\end{equation}
Insertion of $\Phi_k$ from eq. (\ref{kinksolution}) and
evaluation yields
\begin{equation}
E_k = {{2\sqrt{2}}\over 3} {{m^3}\over \lambda}
       \left ( {{N-1}\over {N-3}} \right )
\label{Ek}
\end{equation}
In the limit that $N\rightarrow \infty$, this gives
\begin{equation}
E_k \rightarrow {{2\sqrt{2}}\over 3} {{m^3}\over \lambda}
\label{Eklimit}
\end{equation}
which is precisely the energy of the kink in the $Z_2$
model:
\begin{equation}
L = {1\over 2} (\partial_\mu \phi )^2 - {{m^2}\over 2} \phi^2
     + {\lambda \over 4} \phi^4 + {{m^4}\over {4\lambda}} \ .
\label{Z2model}
\end{equation}

In the large $N$ limit, the kink solution (\ref{kinksolution})
goes to
\begin{equation}
\Phi_k \rightarrow {{m}\over {\sqrt{2\lambda}}}
                    \pmatrix{{\bf 1}_n&{\bf 0}&0\cr
                             {\bf 0}&-{\bf 1}_n&0\cr
                             0&0&-\tanh (mx/\sqrt{2})\cr} \ .
\label{limitofphik}
\end{equation}
Note that this is not traceless because we have discarded a
large number ($N$) of small (order $1/N$) terms.

\section{Perturbative stability}
\label{pertstability}

The procedure for proving perturbative stability is straightforward
though tedious. We consider small deviations $\Psi$ from the 
kink solution as in eq. (\ref{variation}) and
then find the change in the energy due to the perturbations,
\begin{equation}
\delta E [\Psi ] = {\rm Tr} \int dx \Psi \left [ 
    - {1\over 2}{{d^2}\over {dx^2}} + {\bf V}_2 (\Phi_k ) 
                                   \right ] \Psi
\label{pertenergy}
\end{equation}
where the matrix ${\bf V}_2$ is obtained by expanding the potential 
$V$ up to quadratic order in $\Psi$. If the Schrodinger equation
\begin{equation}
\left [ - {{d^2}\over {dx^2}} + {\bf V}_2 (\Phi_k )\right ]  \Psi 
              = \omega \Psi
\label{schrodeqn}
\end{equation}
does not have any negative eigenvalues -- {\it i.e.} regular
solutions where $\Psi \rightarrow 0$ at spatial infinity only 
exist for $\omega \ge 0$ --  then the solution $\Phi_k$ is
perturbatively stable.

The tedious part of this calculation is the evaluation of
${\bf V}_2 (\Phi_k )$. Note that ${\bf V}_2$ is an 
$(N^2-1)\times (N^2-1)$
matrix since there are $N^2-1$ components of $\Psi$, one for
each generator of $SU(N)$. Let us write 
\begin{equation}
\Psi = \sum_{a=1}^{N^2-1} \psi^a T^a
\label{Psicomponents}
\end{equation}
and the elements of ${\bf V}_2$ as $V_{2ab}$. 

Let us discuss off-diagonal perturbations first.
These fall into 5 types. The first type is when
\begin{equation}
T^a_1 \propto \pmatrix{ {\bf \rho}_n&{\bf 0}&0\cr
                       {\bf 0}&{\bf 0}&0\cr
                       0&0&0\cr }
\label{type1}
\end{equation}
where ${\bf \rho}_n$ denotes a non-trivial,
off-diagonal $n\times n$ matrix
and ${\bf 0}$ the trivial $n\times n$ matrix.
The second to fifth type of perturbations are:
\begin{equation}
T^a_2 \propto \pmatrix{ {\bf 0}&{\bf \rho}_n&0\cr
                       {\bf \rho}_n ^{\dag}&{\bf 0}&0\cr
                       0&0&0\cr }
\label{type2}
\end{equation}
\begin{equation}
T^a_3 \propto \pmatrix{ {\bf 0}&{\bf 0}&c\cr
                       {\bf 0}&{\bf 0}&0\cr
                       c^{\dag}&0&0\cr }
\label{type3}
\end{equation}
\begin{equation}
T^a_4 \propto \pmatrix{ {\bf 0}&{\bf 0}&0\cr
                       {\bf 0}&{\bf \rho}_n&0\cr
                       0&0&0\cr }
\label{type4}
\end{equation}
\begin{equation}
T^a_5 \propto \pmatrix{ {\bf 0}&{\bf 0}&0\cr
                       {\bf 0}&{\bf 0}&c\cr
                       0&c^{\dag}&0\cr }
\label{type5}
\end{equation}
where $c$ is a non-vanishing complex $n$-dimensional multiplet.
Then the elements of ${\bf V}_2$ for Type 1 and Type 4 perturbations 
are:
\begin{equation}
V_{2aa} = m^2 {{N \pm 3F}\over {N-3}}  
\label{V2aa14}
\end{equation}
and for type 3 (minus signs) and 5 (plus signs) are:
\begin{equation}
V_{2aa} = \pm {{m^2} \over 2} F (1\pm F) 
\label{V2aa35}
\end{equation}
where the function $F(x)$ is defined in eq. (\ref{Fdefn}).
The elements of ${\bf V}_2$ vanish for type 2 perturbations and
there is no mixing between different off-diagonal perturbations 
either.

The potential $V_{2aa}$ for type 1 and 4 perturbations 
is everywhere non-negative for $N> 3$ and hence the
Schrodinger equation will not have any bound states. The
situation for type 3 and 5 perturbations is less clear
since $\pm F(1\pm F)$ can have either sign. However the
Schrodinger equation has the $\omega =0$ solution
$\psi = 1\mp F$ and this solution has no nodes {\it i.e.}
$\psi \ne 0$ except at infinity\footnote{This zero mode will
be of special interest in Sec. \ref{kinkspace}.}. 
The zero mode solution does not
go to zero at either $x=- \infty$ or at $x=+\infty$.
Then, in order for a solution to go to zero at both spatial 
infinities, $\omega$ has to be chosen to be greater than zero. 
The reason is that the $\omega =0$ solution needs to have smaller
curvature ({\it i.e.} the curvature needs to be more
negative) so that it can vanish at infinity. The curvature
is proportional to $-\omega$ in the region where the potential
is small. Therefore $\omega$ has to be larger. Hence, once again,
there are no bound states to the Schrodinger equation. This 
shows that the solution is stable to off-diagonal perturbations.

Next we consider diagonal perturbations.
These can be classified in three types.
If we write:
\begin{equation}
T_a = \pmatrix{ {\bf R}_a&{\bf 0}&0\cr
                {\bf 0}&{\bf S}_a&0\cr
                  0&0&Q_a\cr}
\label{Tadiagonal}
\end{equation}
where ${\bf R}_a$ and ${\bf S}_a$ are $n\times n$, diagonal
matrices, then the three types of $T_a$ are: 
(1) ${\bf R}_a \ne 0$, ${\bf S}_a=0$, $Q_a =0$,
(2) ${\bf R}_a \ne 0$, ${\bf S}_a \ne 0$, $Q_a =0$, and,
(3) ${\bf R}_a \ne 0$, ${\bf S}_a\ne 0$, $Q_a \ne 0$.
Then there are six cases to be treated in finding the
elements of ${\bf V}_2$.

If both $T_a$ and $T_b$ are type (1), we find
\begin{equation}
V_{2ab} = m^2 {{N+3F}\over {N-3}} \delta_{ab} > 0
\label{V2abbothtype1}
\end{equation}
where we note that we are considering $N> 3$. Since
$V_{2ab}$ is everywhere non-negative, the Schrodinger
equation has no bound states and the solution is stable
to these perturbations.

If both $T_a$ and $T_b$ are type (3) -- there is only
one generator of type (3) -- we find
\begin{equation}
V_{2aa} = {{m^2}\over 2} (3 F^2 -1) \ . 
\label{V2abbothtype3}
\end{equation}
Now the Schrodinger equation is precisely that 
obtained when considering fluctuations about a
$Z_2$ kink \cite{Raj89}. The complete eigenspectrum of 
this equation is known \cite{MorFes54} and the lowest
eigenvalue is $\omega =0$ corresponding to the zero
mode which describes translations of the kink. So there
are no bound states and no instability to these perturbations.

The only other non-trivial perturbations are when $T_a$
and $T_b$ are both type (2). For these perturbations, we
have
\begin{equation}
{\bf R}_a = \alpha_a {\bf 1}_n
\label{Ratype2}
\end{equation}
where $\alpha_a$ is determined from the normalization
of $T_a$. After some algebra, we find
\begin{eqnarray}
\psi_a V_{2ab} \psi_b = {{m^2}\over {N-3}} && [\  
                         (N-3F) \psi_\perp^2 + \{ (N-3F) - 
                                                  \nonumber \\
  &&6 \sigma^2 \, (N-1) (1-F) \} \psi_\parallel ^2 \ ]
\label{psiaV2abpsib}
\end{eqnarray}
where, we have defined the unit vector $\hat \alpha_b$,
decomposed $\psi_a$ parallel 
($\psi_\parallel = {\hat \alpha}_a \psi_a$)
and perpendicular ($\psi_\perp$) to the vector $\alpha_a$,
and written
$$
\sigma^2  \equiv \sum_a \alpha_a \alpha_a  \ .
$$
The coefficient of $\psi_\perp ^2$ is positive since
$N > 3$ and hence this perturbation cannot cause an
instability. The coefficient of $\psi_\parallel^2$ needs
to be checked.

Using the normalization of $T_a$, we obtain
\begin{equation}
\sigma^2 = {1\over {2(N-1)}} 
\label{alphaaalphaa}
\end{equation}
Inserting this into eq. (\ref{psiaV2abpsib}), we find
that the coefficient of $\psi_\parallel^2$ is simply
$+m^2$. Hence there is no instability to
diagonal type (2) perturbations.

This explicit analysis shows that the solution is 
stable to all perturbations but this does not imply
that the solution minimizes the energy globally.

\section{Gauge fields}
\label{gaugefields}

The inclusion of gauge fields, $A_\mu = A_\mu^a T^a$, has no
effect on the existence of the static solution in
eq. (\ref{kinksolution}). This can be seen by noting that the 
terms in the energy that involve the gauge fields are:
${\rm Tr}({\bf E}^2 + {\bf B}^2)$,
$-{\rm Tr}([A_x , \Phi ]^2)$ and
$i{\rm Tr}(A_x [\Phi , \partial_x\Phi]$. The first two terms
are quadratic in the gauge fields while the last one vanishes
because $[\Phi , \partial_x \Phi ] =0$ for the solution.
Hence $A_\mu =0$ is a solution of the equations of motion.

The presence of gauge fields does not have any effect on
the perturbative stability of the solution. To see this
one can check that both the quadratic order terms in the gauge
field are non-negative. For the first term this is explicit
while for the second term one uses the fact
$[A_x , \Phi ] = - [A_x , \Phi ]^\dagger$ since $A_x$ and
$\Phi$ are Hermitian.  Hence the kink solution with $A_\mu=0$
is stable to perturbations in the gauge fields.

\section{Space of kinks}
\label{kinkspace}

Different boundary conditions will, in general, lead to
different kink solutions. Therefore kinks with different
boundary conditions fall into different classes -- kinks 
belonging to different classes cannot be transformed into 
one another by global $SU(N)$ rotations. 
Here we would like to find the degeneracy of a kink
solution {\it i.e.} the space of boundary conditions that
lead to degenerate kink solutions. There is clearly an $SU(N)$
global degeneracy but this is not very interesting since it
applies to any field configuration in the theory. It is of
greater interest to only consider those global $SU(N)$ 
transformations that leave $\Phi_- \equiv \Phi( x=-\infty)$ 
unchanged but act non-trivially on 
$\Phi_+ \equiv \Phi( x=+\infty)$. If we denote the unbroken
symmetry groups at $x=\pm \infty$ by $H_\pm$, such transformations
belong to $H_-$. But the transformations that belong to 
$K\equiv H_+ \cap H_-$ will act
trivially on both $\Phi_-$ and on $\Phi_+$. Therefore the
space of boundary conditions at $x=+\infty$ leading to 
degenerate kinks is given by $H_-/K$.

In addition to the degeneracy due to different boundary
conditions, any kink solution will have an ``internal''
symmetry group, denoted by $I$. This group will contain
all those transformations that leave unchanged the whole 
kink solution (including the boundary conditions). 
So we have $I \subseteq K$ and the space of degenerate kinks 
is $H_-/I$.

The kink classification problem can be described in more 
detail as follows. 
Suppose we fix the boundary condition at $x=-\infty$ to be 
$\Phi_-$, then the only constraint on the boundary condition
at $x=+\infty$ is that it should be in the distinct topological
sector. There is a full vacuum manifold (mod $Z_2$) worth of 
choices for $\Phi(x=+\infty ) \equiv \Phi_+$. For 
certain choices of $\Phi_+$ we can solve the equations of 
motion and obtain a 
set (described by the space $K/I$) of kink solutions that 
extremize the energy. Let the value of this energy 
be $U[\Phi_+; \Phi_- ]$ where we have explicitly indicated
that different choices of boundary conditions can lead to kink
solutions of different energy. We are interested in the 
space of minima of the ``potential'' $U[\Phi_+; \Phi_-]$ 
with respect to $\Phi_+$. The global minima
of this potential will describe the lowest energy kink solutions
in the model and may be termed the ``kink vacuum manifold''.
Other local minima will describe kink solutions that are
separated from the lightest kink by an energy barrier. In
this way it might be possible to obtain ``generations'' of
stable kinks having the same topological charge but differing 
in their energies.

The existence of the kink solution found in the previous sections
does not tell anything about whether it is a minimum of 
$U[\Phi_+; \Phi_-]$. To examine if the solution is a local
minimum we can do a perturbative analysis where the perturbations
are required to vanish at $x=-\infty$ -- so as to hold $\Phi_-$
fixed -- but are not required to vanish at $x=+\infty$ -- since we
want to find changes in the energy when the boundary conditions
are varied. Going back to Sec. \ref{pertstability} we find that
all but one of the non-trivial Schrodinger potentials are positive 
(for $N > 3$) at both spatial infinities. This means that a 
perturbation that does not vanish at both spatial infinities, 
gives a divergent contribution to the energy. Hence the kink
solution is stable under these perturbations of the boundary
conditions. The only exceptional case is the off-diagonal 
perturbation of the type in eq. (\ref{type3}). The potential
(eq. (\ref{V2aa35})) goes to a positive value at $x=-\infty$
but vanishes from below at $x=+\infty$. In fact, as described
below eq. (\ref{V2aa35}), there is a zero mode for this
perturbation that vanishes at $x=-\infty$ but goes to a
non-vanishing constant at $x=+\infty$. So this mode is a
``dangerous'' one and needs to be examined further.

A closer inspection of this mode shows that it corresponds
to gauge rotations of the field $\Phi$ which leave $\Phi_-$
invariant but rotate $\Phi_+$. In other words, the zero mode
rotates the kink within its own class described by the space
$H/I$. Hence, the perturbation under consideration is not
an instability but a gauge rotation. Therefore the class of 
kinks that has been found describes a set of local minima of 
$U[\Phi_+ ;\Phi_-]$.

There is a subtlety in the discussion above which we have
glossed over. To obtain the Schrodinger equation in eq.
(\ref{pertenergy}) we have to perform an integration by
parts and assume that the boundary contributions vanish.
However, here we are considering perturbations that do
not vanish at infinity. This fact means that there is an
extra contribution to $\delta E_{\infty}$ in eq. (\ref{pertenergy})
given by
\begin{equation}
\delta E_\infty  = {1\over 2} {\rm Tr} \left [ 
  \Psi {{d\Psi} \over {dx}} \right ]_{-\infty}^{+\infty}
\label{bcenergy}
\end{equation}
The contribution at $x=-\infty$ vanishes because we
are choosing $\Psi (-\infty )=0$ but the contribution
at $x=+\infty$ does not obviously vanish and depends
on the derivative of the perturbation at infinity.
However, since we are only interested in finite energy
field configurations and the energy density contains
the term ${\rm Tr} (\partial_x \Psi )^2$, we must require 
that the derivative of the perturbations vanish at spatial 
infinity. So it is rigorous to take $\delta E_\infty =0$
even if the perturbations do not vanish at infinity.

A more extensive discussion of the kink classification problem
is left for future work.

\section{Summary}
\label{summary}

We have considered $SU(N)\times Z_2$ models. On the
basis of topology, the model will contain topological kink 
solutions. A general technique for constructing the
solutions is not known. Here, by using some guesses
and by choosing a special relation between parameters
(eq. (\ref{paramchoice})),
we have analytically constructed a class of kink solutions 
(see eq. (\ref{kinksolution})). The energy of the solutions 
is 
\begin{equation}
E = {{2\sqrt{2}}\over 3} \left ( {{N-1}\over {N-3}} \right )
                {{m^3}\over {\lambda}}
\end{equation}
The limiting value for large $N$ is equal to the
energy of a $Z_2$ kink with mass parameter $m$ and 
coupling constant $\lambda$.
We have explicitly checked that these kink solutions are 
perturbatively stable. It is not known if the solutions 
are globally stable and this remains an interesting open
problem. 

We have then described the space of kinks as partitioning
into distinct classes. All members have the same topology, yet
elements of different classes are not expected to have the same 
energy.  The solutions constructed above describe only one of the
(unknown number of) classes of kinks and might lie on 
the ``kink vacuum manifold'' -- the manifold consisting of
the least energetic kinks in the model.

We hope that the solutions found here can be used as a 
guide to the construction of other topological defect 
solutions in complicated field theories.

\acknowledgements

I am grateful to Levon Pogosian for discussions and 
especially his input in Sec. \ref{kinkspace}. I also
thank Sandip Trivedi for discussions.
This work was supported by the DoE.

\end{document}